\documentclass[11pt]{article}

\usepackage{amsmath,amssymb,amsfonts,mathrsfs,amsthm,mathtools}

\usepackage{sgame}

\usepackage{graphicx,epsfig,subfigure,float}
\usepackage[space]{grffile}
\usepackage{color}

\usepackage{geometry}
\usepackage[bottom]{footmisc}
\usepackage{setspace}

\usepackage{enumerate}
\usepackage{url}
\usepackage{lscape}
\usepackage{comment}

\usepackage{calligra,dsfont,bbm,gensymb}

\usepackage[round]{natbib}
\usepackage{authblk}
\usepackage{hyperref} % load last

\usepackage{pdfpages,lineno,lipsum,accents}

\title{Conditional Recall}
\author{Christoph Schlegel and Xinyuan Sun}
\date{}

\begin{document}

%%% The following commands remove the headers in your paper. For final 
%%% papers, these will be inserted during the pagination process.

%%% The next command prints the information defined in the preamble.

\maketitle 

% The year is 2025, Johnson and Johnson has just released a new drug: X. X allows people to choose which memories they want to forget and let people know that they indeed forgot those memories. You are a student at Harvard Law with the dream of changing the law industry for the better, and now your dealer has offered you X at low prices, wdyd?

% \newpage

% \tableofcontents
\begin{abstract}In the neon-lit nights of 2026, Johnson \& Johnson unveiled X. A pill, not larger than a snowflake, that promised a tempest of change. This miraculous drug didn't just allow people to cherry-pick memories to erase from their minds, it could also leave a reminder of this erasure in the minds of those who ingested it.

Amidst the iconic red-bricked walls of Harvard Law, you, with books in one hand and dreams in the other, are on a mission. You are not just another student; you carry the hope of revolutionizing the archaic chambers of the legal world. Each night, as you pore over the tomes of law, you wonder what greatness society can achieve.

On a cold evening, your phone buzzes. It's Dex, your old college friend turned underground dealer. His message is simple: ``Got X. Special price for you.'' The temptation swirls around you. Would you trade the lessons of the past for a clearer, yet incomplete future? The decision rests in your hands.

We explore the game theoretic implications of a technology (such as TEEs) that allows agents to commit to forget information and discuss several applications.
\end{abstract}
\section{Introduction}

Perfect recall is a foundational assumption in extensive form games since~\cite[]{Kuhn1953}. While this assumption naturally models human cognition — as humans cannot selectively forget information — it may be unnecessarily restrictive for artificial agents. In particular, Large Language Models (LLMs) can implement conditional recall through context segmentation: multiple instances can each hold private information, verify each other's reasoning chains for information leakage, and commit to forgetting the interaction. Credibility of commitments to forget can be enforced, for example, through the use of Trusted Execution Environments (TEEs). These capabilities suggests new mechanism design possibilities.

We formalize this insight through a game-theoretic framework in which players can choose to forget information they have previously learned while maintaining time consistency. Using a hypothetical ``pill X'' as our primitive, we identify several classes of games where conditional recall enables more efficient equilibria: (i) information markets where Arrow's paradox previously prevented trade, (ii) information sharing schemes where one-time use of information is enforced, (iii) bargaining protocols where reputation effects would otherwise sustain inefficient outcomes, and (iv) Coasian dynamics where the seller can extract more value by being able to credibly forget. 

Through a series of strategic scenarios — from corporate negotiations to military intelligence — we demonstrate how conditional recall mechanisms strictly dominate classical solutions that assume perfect recall. Our results suggest that as artificial agents increasingly participate in strategic interactions, mechanism designers should exploit their ability to selectively forget information, a capability that transcends human cognitive constraints.
\section{Background and Related Work}

\subsection{Trusted Execution Environments and Confidential Computing}

Trusted Execution Environments (TEEs) are secure computing environments that provide strong isolation guarantees through hardware-based security mechanisms. TEEs enable the execution of sensitive code in a protected enclave, ensuring confidentiality and integrity of both code and data, even in the presence of a compromised operating system or malicious host~\citep{mckeen2013innovative, cheng2024intel, costan2016intel}. And confidential computing capabilities on NVIDIA Hopper GPUs show minimal impact on performance (within 7\%), making confidential inference of LLMs practical~\citep{zhu2024confidential, anthropic2025confidential}. This practical availability is crucial for deploying the types of AI-mediated negotiation and information disclosure mechanisms we study in this paper.

\paragraph{Hardware Implementations.} Modern TEE implementations include Intel TDX (Trusted Domain eXtentions)~\citep{cheng2024intel}, intel SGX (Software Guard Extensions)~\citep{costan2016intel}, ARM TrustZone~\citep{pinto2019demystifying}, and AMD SEV (Secure Encrypted Virtualization)~\citep{kaplan2016amd}. These technologies create isolated execution environments where sensitive computations can occur without exposure to the host system, allowing external parties to cryptographically verify that specific code is running in a genuine enclave~\citep{anati2013innovative}. 

\paragraph{Security Properties and Limitations.} TEEs provide three critical security properties: (1) \textit{confidentiality} - ensuring that data within the enclave cannot be read by unauthorized parties, including the host OS; (2) \textit{integrity} - guaranteeing that code and data cannot be tampered with during execution; and (3) \textit{remote attestation} - enabling verifiable proof that specific code is running in a genuine TEE~\citep{sabt2015trusted}. However, TEEs are not immune to attacks. Side-channel attacks exploiting cache timing~\citep{brasser2017software}, speculative execution~\citep{van2018foreshadow}, and memory access patterns~\citep{shinde2017preventing} have been demonstrated. Oblivious RAM (ORAM)~\citep{goldreich1996software} can mitigate memory access pattern leakage, while careful implementation practices can reduce other attack surfaces.

\paragraph{Confidential Inference and AI Systems.} The deployment of machine learning models raises significant privacy concerns for both model providers (who want to protect intellectual property) and users (who want to protect sensitive input data)~\citep{tramer2018slalom, ohrimenko2016oblivious}. TEEs provide a practical approach to confidential inference by executing model inference within a secure enclave~\citep{anthropic2025confidential}. The model parameters can be loaded into the enclave, encrypted user inputs can be decrypted inside the TEE, inference performed, and only the result returned - all while ensuring that neither the cloud provider nor potential attackers can access the model or data. Remote attestation allows users to cryptographically verify the model's hash and the TEE's security properties before providing sensitive inputs, enabling accountability and auditability of AI systems in high-stakes applications~\citep{anthropic2025confidential}. More recent work has explored TEEs for implementing one-time programs~\citep{goldwasser2008one, zhao2019one, eldridge2022one}, which allow code to execute exactly once with verifiable deletion of secrets afterward - a critical primitive for credible forgetting mechanisms.

\subsection{Coordination Over Shared Private State}

Many coordination problems require parties to share private information to achieve efficient outcomes, yet such sharing creates risks of exploitation. Arrow~\citep{arrow1972economic} identified this tension in information markets: buyers want to verify information quality before purchasing, but inspection reveals the information itself, eliminating the seller's ability to capture value. This \textit{disclosure problem} exemplifies a broader class of coordination failures where parties cannot credibly commit to using shared information only for mutually beneficial purposes~\citep{anton2002sale}.

Recent work demonstrates that AI agents can reduce information asymmetries in markets~\citep{rahaman2024language}, but fundamental commitment problems remain: how can parties ensure that information shared for evaluation or negotiation will not be exploited? Traditional solutions rely on legal instruments like patents, NDAs, and trade secret law, but these impose high transaction costs and suffer from imperfect enforcement~\citep{graham2009small, contigiani2019trade}. Even sophisticated approaches like NDAI agreements—NDAs specifically designed for AI-mediated interactions—face challenges in ensuring compliance and preventing information leakage~\citep{stephenson2025ndai}. Patents require public disclosure that enables competitors to design around them. NDAs require costly monitoring and ex post litigation. Trade secrets limit collaboration. These frictions lead to market failures: innovations may not be developed~\citep{dushnitsky2009limitations}, value-creating collaborations fail to form~\citep{nelson1959simple}, and resources are misallocated to protecting rather than creating information.

The disclosure problem is closely related to hold-up in incomplete contracts~\citep{hart1988incomplete, grossman1986costs}: parties making relationship-specific investments face the risk that others will exploit sunk costs in subsequent bargaining~\citep{aghion1987contracts}. More broadly, any coordination requiring shared private state faces a fundamental commitment problem: parties cannot credibly commit to not exploiting information once revealed—a form of dynamic inconsistency that reduces ex ante investment and trade. This problem becomes particularly acute in multi-agent AI systems, where information can be processed and exploited at scale.

\subsection{Commitment Devices and Strategic Coordination}

Commitment devices enable agents to credibly constrain future actions, often improving equilibrium outcomes. Schelling~\citep{schelling1960strategy} showed how limiting one's options can be strategically advantageous. However, commitment is only valuable if credible—parties must believe it will be honored~\citep{kalai2010commitment}.

Traditional commitment mechanisms face significant challenges. Legal contracts suffer from incomplete specification and costly enforcement~\citep{hart1988incomplete}. Reputation mechanisms require accurate monitoring and patient players~\citep{kreps1982reputation}, and can sustain inefficient delays in bargaining~\citep{abreu2000bargaining}. Critically, mediators or mechanism designers may themselves behave strategically: Akbarpour and Li~\citep{akbarpour2020credible} show that rational auctioneers who can exploit information make many standard mechanisms infeasible without additional commitment devices.

Tennenholtz~\citep{tennenholtz2004program} introduced program equilibrium, where agents commit to source code and inspect each other's programs before interaction. Recent work explores commitments to learning algorithms~\citep{oesterheld2022similarity} and cryptographic commitments to strategies~\citep{ferreira2020credible}. However, these approaches face challenges when code cannot be fully inspected, can be modified after inspection, or when imperfect information prevents complete verification.

\subsection{AI Agents and Imperfect Recall}

AI agents are increasingly deployed in high-stakes economic settings where coordination failures have severe consequences—from algorithmic trading~\citep{budish2015high, easley2011microstructure} to automated auctions~\citep{mehta2023auctions, banchio2022artificial, kolumbus2022auctions} to dynamic pricing~\citep{assad2020algorithmic, chen2016empirical, wieting2021algorithms}. Foundation models, when augmented with tool use and memory~\citep{wang2023voyager}, demonstrate capabilities essential for strategic interaction: reasoning about other agents' beliefs, simulating negotiations, and evaluating complex agreements.

Classical game theory assumes perfect recall~\citep{Kuhn1953}—players remember all past actions and observations. This accurately models human cognition but may be unnecessarily restrictive for artificial agents. Relaxing this assumption reveals strategic possibilities: absent-mindedness~\citep{piccione1997interpretation} creates dynamic inconsistency, while selective forgetting can eliminate inefficient reputation effects. Recent work asks whether forgetting can be strategically designed and committed to~\citep{conitzer2019designing}.

\paragraph{Credible Forgetting as Commitment.} If an agent can credibly commit to forget certain information, this enables new coordination mechanisms. An agent who credibly forgets past offers cannot sustain reputation-based threats, potentially improving bargaining efficiency~\citep{abreu2000bargaining}. An auctioneer who credibly forgets losing bids cannot exploit bid information~\citep{akbarpour2020credible}. The challenge is making forgetting credible: human agents cannot verifiably forget, and computational agents could maintain hidden copies unless constrained by technology.

One-time programs (OTPs)~\citep{goldwasser2008one} formalize credible forgetting: computations that execute exactly once with all secrets provably deleted. While theoretically achievable through quantum mechanics~\citep{broadbent2013quantum}, practical implementations require TEEs~\citep{zhao2019one, eldridge2022one}. OTPs provide commitment to future ignorance—the program cannot be re-executed, and attempts to do so are detectable.

\paragraph{AI Agents in TEEs.} The combination of confidential inference and conditional recall creates powerful capabilities for coordination. Agent loops in TEEs can process sensitive information while remote attestation ensures specified protocols are followed. Crucially, TEEs enforce that after evaluation, sensitive information is deleted and cannot be recovered—even by parties who provided it. This enables:
\begin{itemize}
    \item Information disclosure without expropriation: sellers can reveal innovations for evaluation, knowing information is deleted if no agreement is reached
    \item Efficient bargaining without reputation effects: parties can share valuations, knowing this information cannot build reputations causing delay  
    \item Privacy-preserving mechanism design: agents participate in complex mechanisms with guarantees about information use and deletion
\end{itemize}

The practical deployment of autonomous AI agents in TEEs~\citep{anthropic2025confidential,hu2025trustless}, combined with the theoretical framework of conditional recall, suggests that AI-mediated coordination with credible forgetting is increasingly practical.

\subsection{Related Economic Theory}

Our analysis connects to several economic literatures. In Coasian dynamics~\citep{coase1972durability}, monopolists face dynamic inconsistency: they want to lower prices after selling to high-valuation buyers, but buyers anticipate this, reducing willingness to pay. This directly parallels information sellers who might resell to others at lower prices. Credible commitment to not making future sales—or to forgetting buyer identities—can restore monopoly power.

In disclosure games~\citep{grossman1981informational, milgrom1981good}, unraveling results show even bad types may be forced to disclose. However, these assume information cannot be undisclosed. Conditional recall mechanisms enabling "disclosure with deletion" may significantly alter equilibrium disclosure strategies. In repeated bargaining~\citep{abreu2000bargaining, fudenberg1991game}, reputation effects can sustain inefficient delays. If players can credibly forget past offers, certain reputation strategies become infeasible, potentially improving efficiency.

\section{Model and Framing}
We consider extensive form games of incomplete information. Such game $\Gamma=(N,A,\mathcal{H},\mathcal{I},u)$ consists\footnote{Subsequently, we follow the definitions of \cite{Rubinstein2}.} of a finite set of players $N$ (including a player called "nature"), a set of {\bf actions} $A$, a set $\mathcal{H}$ of sequences of actions, called {\bf histories}, such that if $a_1,\ldots,a_K$ is a history and $L<K$ then $a_1,\ldots,a_L$ is also a history,\footnote{We subsequently say that history $a_1,\ldots,a_K$  is an {\bf extension} of history $a_1,\ldots,a_L$} payoff functions $u_i:Z\rightarrow \mathbb{R}$ for each $i\in N$ where $Z$ is the set of terminal histories (those histories that cannot be further extended), a mapping from non-terminal histories to players and a mapping from non-terminal histories to actions which prescribe which player takes an action after the history and which actions are available to them (denote by $\mathcal{H}_i$ for each $i\in N$ the set of histories after which $i$ takes an action and $A(h)\subseteq A$ to be the set of available actions after history $h$) and a partition of $\mathcal{H}_i$ for each $i$ into {\bf information sets} $\mathcal{H}_i=\bigcup_{I\in\mathcal{I}_i}I$ such that the set of available actions are the same after each history in the same information set. Information sets model which states of the world are indistinguishable for the players when choosing an action, based on the information that they take into consideration when taking an action. In the following, for $h,h'\in\mathcal{H}$ we write $h'\succ h$ if $h'$ is an extension of $h$ and for $I,J\in\mathcal{I}_i$ we write $J>_i I$ whenever there is a $h\in I$ and a $h'\in J$ with $h\succ h'$. Note that $\succ$ is a strict partial order, i.e. it is transitive and irreflexive, whereas $>_i$ is not necessarily a strict partial order. 

What does it mean to forget in an extensive form game? For a given history $h$ and player $i$ we can consider the sequence $X_i(h)$ of information sets he has encountered and the actions he has chosen along the history $h$. We say that player $i$ has {\bf Perfect Recall} if $X_i(h)=X_i(h')$ whenever $h$ and $h'$ are in the same information set. Thus, if two histories are in the same information set for player $i$, he couldn't distinguish between them even if he takes into consideration his entire experience of playing the game up to that information set. Perfect recall is equivalent (see~\cite{ritzberger1999recall}) to the simultaneous fulfillment of three
independent properties: Players never forget what they did; they never forget
what they knew;\footnote{Player $i$ remembers what he previously knew if for information sets $I,J\in\mathcal{I}_i$ with $I\succ J$ we have $\{h\in T:h\succ I\}\subset \{h\in T: h\succ J\}$, i.e. if a history is already excluded at a past information set then so it is at a subsequent information set. Player $i$ remembers his past actions if we cannot further partition $\mathcal{I}$ by only relying on his own past choices.
} and (for a given player $i$), past, present, and future have
unambiguous meaning. The last requirement means that for $i$ the relation $>_i$ is a strict partial order.  If $>_i$ is transitive\footnote{In-transitivity is not far-fetched as it seems: Sending transactions to a blockchain mempool is an example of a situation with ambiguity about future, past and present, as we do not know how nature (searchers, builders and proposer) orders transactions.} but not irreflexive this is commonly called {\bf absent-mindedness},\footnote{Absent-mindedness has interesting paradoxical implication, see˙\cite{Rubinstein}} i.e. we are in a situation where $i$ does not remember whether he already took an action or not. In the following, X is a technology that allows to conditionally forget or being absent-minded (which can be interpreted as another type of forgetting) while maintaining the requirement that $>_i$ is transitive. X can be customized to choose those actions that the player wants to forget and those that he wants to remember. We assume that $X$ can only be used once.

Formally, if player $i_0$ has access to $X$ for game  $\Gamma$ with perfect recall this transforms the game into a new game $\Gamma^X$ where prior to choosing an action player $i_0$ may choose the action to take $X$, and taking $X$ has an effect on the information sets that $i_0$ subsequently traverses.
The set of histories is \begin{align*}\mathcal{H}^X=\mathcal{H}\cup\{(a_1,\ldots,a_{K-1},X,a_K,\ldots, a_L):\\(a_1,\ldots,a_K)\in\mathcal{H}_i,(a_1,\ldots,a_L)\in\mathcal{H},K<L\},\end{align*}
utility does not depend on taking $X$, i.e. $u_i^X(h)=u_i^X(h\setminus X)$, 
and the information set partition $\mathcal{I}^{X}_{i_0}$ for player $i_0$ in $\Gamma^X$ has the following properties: 
\begin{enumerate}
\item Histories in which $i_0$ doesn't take $X$ are partitioned as before and distinguishable from histories in which he does take $X$,  i.e. $\mathcal{I}_{i_0}\subset\mathcal{I}^{X}_{i_0}$, (not taking $X$ doesn't modify information sets and the player remembers whether he took $X$ or not),
\item If $h,h'\in\mathcal{H}_{i_0}^X$ are in different information sets, then $h\setminus X,h'\setminus X\in\mathcal{H}_{i_0}$ are in different information sets in the original game (if $i_0$ remembers something after taking $X$ he also remembers it in the original game, taking $X$ leads to a coarsening of information sets),
%\item Information sets can be partially ordered so that any extension of a history is in an information set that succeeds the information set of the history (so that there is no ambiguity about past present and future). 
\item The ordering $>^X_{i_0}$ on information sets is transitive.
\item For histories $h,h'\in \mathcal{H}_{i_0}^X$ that are  extended in the same way $(h,a_1,\ldots,a_K),(h',a_1,\ldots,a_K)\in\mathcal{H}_{i_0}^X$, if $h$ and $h'$ are in the same information set and $A(h,a_1,\ldots,a_K)=A(h',a_1,\ldots,a_K)$ then $(h,a_1,\ldots,a_K)$ and $(h',a_1,\ldots,a_K)$ are in the same information set (if $i_0$ forgets something he cannot remember it later)
\item any histories $h,h'\in \mathcal{H}_0^X$ in which $i_0$ takes $X$ and that only differ in when $X$ is taken, $h\setminus X=h'\setminus X$, are in the same information set (the effect of $X$ is independent of when it is taken),
\end{enumerate}
whereas other players observe $i_0$ taking $X$ so that their information set partitions  are:
$$\mathcal{I}_j^X=\mathcal{I}_j\cup\{\{h\in\mathcal{H}_j^X\setminus\mathcal{H}_j:h\setminus X\in I\}:I\in\mathcal{I}_j\}.$$

\begin{figure}[t]
    \centering
    \includegraphics[scale=0.7]{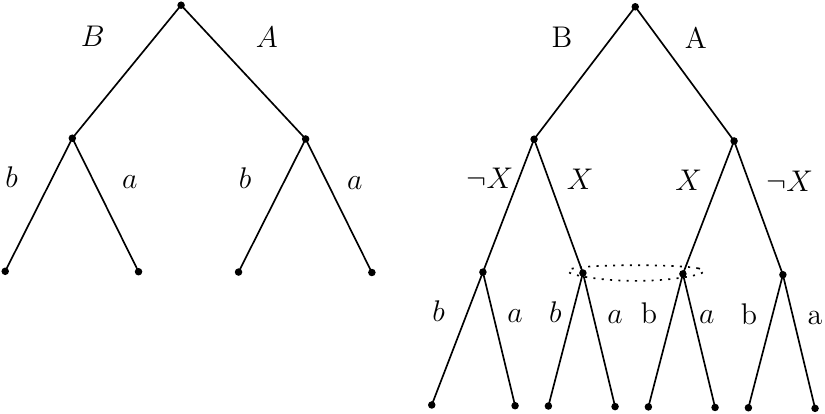}
    \caption{The original game (left): player $1$ chooses between actions A and B and player between actions $a$ and  $b$. In the modified game where the second player has access to $X$ (right),  the second player can choose to take $X$ which makes him forget whether the first action taken is $A$ or $B$.}
    \label{Untitled}
\end{figure} 
Note that depending on the version of $X$, it allows to forget actions after having observed them, as well as committing to forget future observations. The fourth item of our definition is for the sake of conceptual simplification: it is satisfied in all of our subsequent examples, but it restricts the powers of X. It would be an interesting extension for future work to extend the power of X by allowing to make the functionality of X be a function of history, or even make several versions of X available that allows to customize its effect during the game.

The definition extends naturally to the case where multiple players have access to $X$ (each of them having their own version of $X$). %In that case we impose that after each non-terminal history each player has simultaneously the choice to take $X$ (so that player do not know whether another player takes $X$ concurrently, when making the choice to take $X$, but know (unless they forget) whether another player took $X$ in the past). 

Consider Figure $1$ for an example of a game and the corresponding game with $X$.

%We investigate whether having X allows Pareto improvements. And whether having X with delegation allows more Pareto improvements.
\section{Applications}
Subsequently we go through a variety of examples to demonstrate the power of $X$.

\subsection*{Arrow's information paradox}

% Information carries value: the knowledge of the list of all CIA agents in Russia is high stakes. Suppose you are a defecting American agent and would like to trade your knowledge of your peer's whereabouts to the KGB, you enter the room, being a trained agent with 20 years of experience, you are confident of handling uncertain and sensitive situations whenever they come up, your reflex is still there, along with the luck goddess. Suddenly somebody puts a gun to your head and asks you to sit down, you did, looking across the table, sitting are two mob-looking bear-fighting Russians, they ask you of the information. A sudden unease began spreading in your mind, you know the works of those KGB dudes, they will not hesitate to kill you to prevent their information being devalued after you telling them. Afterall, information is highly memetic and the more copies there are, the less valuable each copy is. You now have to calculate what information to tell them, do you lie or not, or do you pretend to not know anything and back out the trade. 

You are an US agent in the Soviet Union wishing to defect. You hold the list of all US agent names and would like to trade that list for riches in Moscow. However, knowing the KGB would not pay you once you have told them the names and the KGB would not buy the information without having verified it, you have to calculate: what information to tell them, do you lie or not, or do you pretend to appear uninformed to back out of the trade. 

This situation is an instance of Arrow's information paradox~\cite{arrow1972economic}: the potential buyer of a piece of information wants to inspect its content before deciding if they should buy it. However, once the buyer has inspected the information, they already own it and cannot ``unknow'' the information due to their inability to forget. The seller, knowing all of this, would then choose to either 1) hide the information or 2) use the legal system and a mediator (an IP lawyer or some trusted third party).

\begin{figure}[t]
    \centering
    \includegraphics[scale=0.5,angle=90]{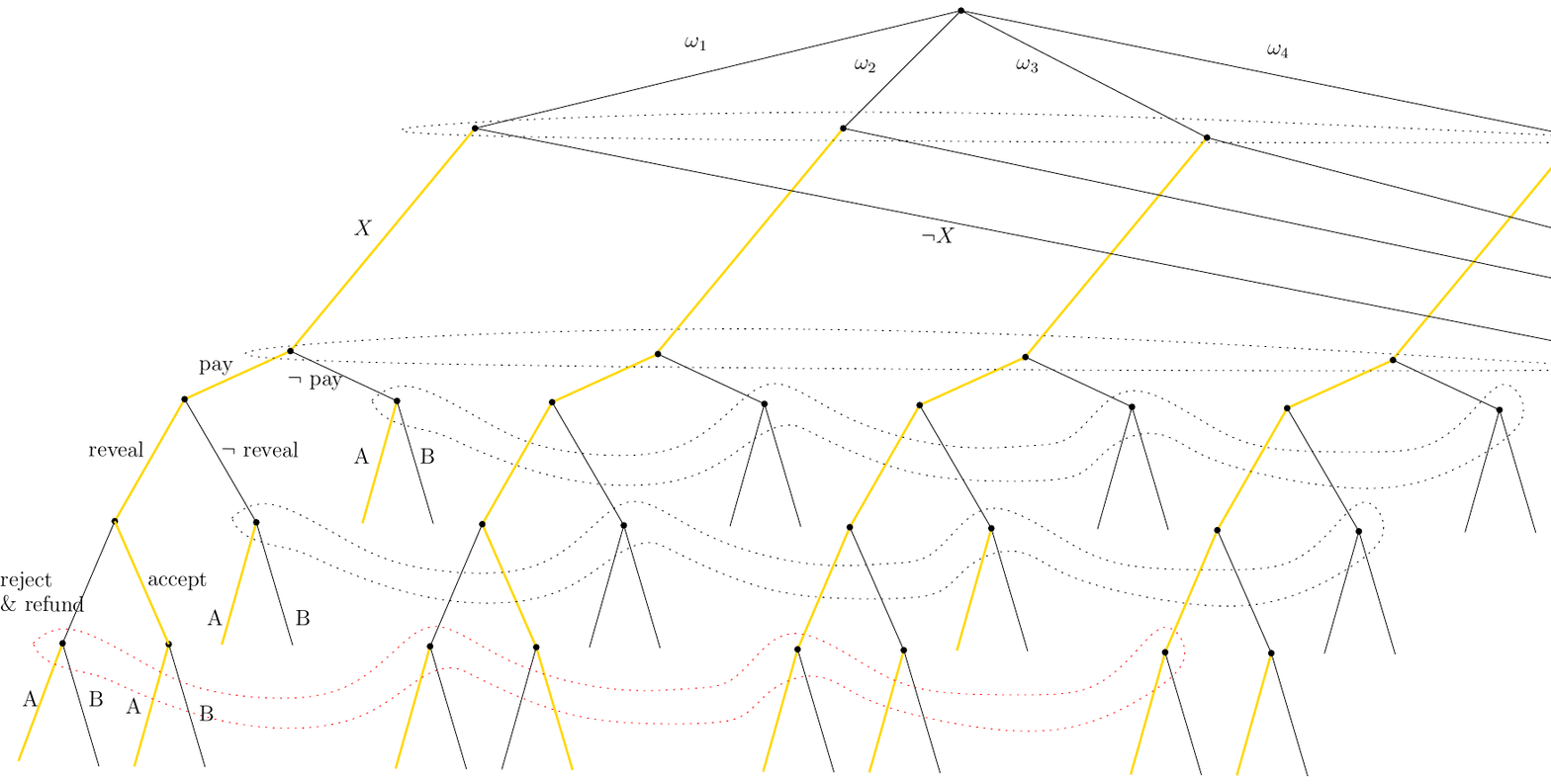}
    \caption{An instantiation of Arrow's information paradox and it's solution with $X$: Nature chooses a state of the world $\omega\in\{\omega_1,\omega_2,\omega_3,\omega_4\}$ and reveals it to player $2.$ Player $1$ needs to take an action A or B. Suppose e.g. if the state of the world is $\omega_1$ then it is optimal for him to choose $A$ and if the state of the world is $\omega_2$ and it is optimal for him to choose $B$. If the state $\omega_3$ or $\omega_4$, it doesn't matter whether he takes action $A$ or $B$. However, knowing whether the state is $\omega_3$ or $\omega_4$ can be relevant to a third party, so that knowing the state has still value (but the same value in either state)  for player $2$.  In the modified game where the second player has access to $X$,  the second player can choose to take $X$ which makes him forget the state that was previously revealed by player $1$ in case that he rejects to buy that information. This introduces the red information set. For appropriate payoffs, the contract has the depicted (highlighted in yellow) subgame perfect equilibrium.}
    \label{Untitled}
\end{figure} 

If they choose to hide the information, then adverse selection will take effect: now buyers are paying the average price for information, and sellers of high-value information will leave the market because they will not get a good deal. As a result, buyers will have lower willingness to pay because the average quality of the information has decreased. This ``market for lemons" situation will persist until the market reaches a point where there is only valueless information and inevitably collapse. 

Using lawyers and IP or patent law, on the other hand, creates high transaction costs, and a lot of the value in the information trade will be extracted by the lawyer and the legal system (assuming lawyers are scarce and they are the only source of credibility and mediation you could use for information transactions). Using lawyers requires, moreover, a functioning legal system to rely on, which might be a bit far-fetched for such a risky spy operation. 
\begin{figure}[t]
    \centering
    \includegraphics[scale=0.5,angle=90]{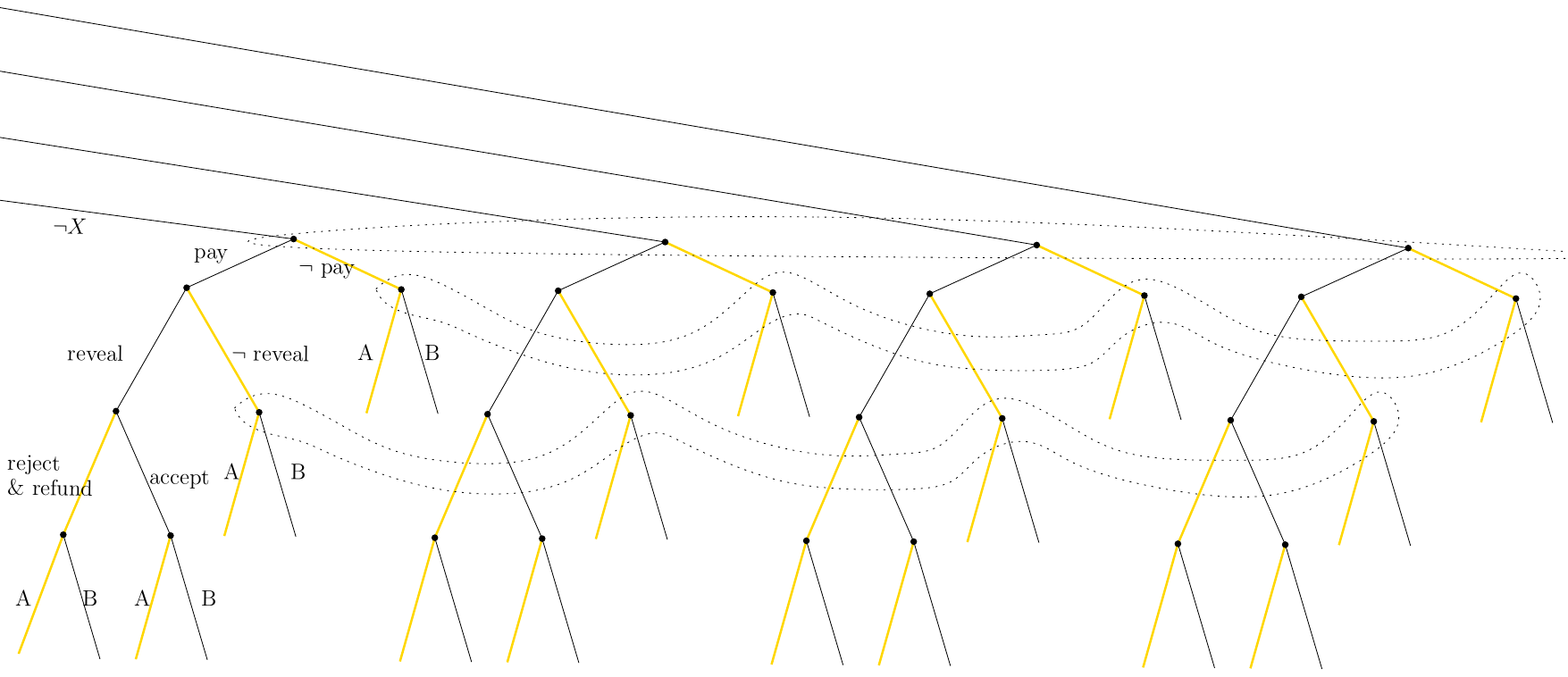}
    \caption{Continuation of the game tree from the previous page.}
    \label{Untitled}
\end{figure} 
To solve the contracting problem you come up with the following scheme (see Figure-2 for a simple instantiation of the contracting problem with $X$): The KGB pays you (e.g. through a third party intermediary or some other escrow scheme) before having verified the information and keeps you in custody while checking the veracity and quality of the information. If they do not like the information, they demand you to organize the money to be returned. You return (e.g. through the intermediary) the money once they have ingested pill $X$ and they release you from custody once they have received the money back. If they like the information, you will keep the money and can leave.
Moreover, the KGB can mandate you to also ingest pill X, ensuring the information is purged from your memory. This mechanism mitigates the risk of subsequent unauthorized disclosures, which would otherwise devalue the information for the buyers. Consequently, pill X serves as a novel commitment device, potentially increasing the efficiency and reliability of high-stakes information markets.

\subsection*{Forgetting as substitute for commitment}
Forgetting can often be used as a substitute for commitment. There are many examples of situations where an player may change their mind about an action once they receive new information, even though they would be better off when able to commit to a course of actions (which actions to take conditional on the content of the information) prior to receiving the information.\subsubsection*{The forgetful monopolist}
The following situation is a version of Coasian dynamics, where forgetting works as a substitute of price commitment.\footnote{This particular version of the example is due to Wei Dai's: http://www.weidai.com/monopoly-memory.txt} This is a problem faced, for example, in online sales where a seller might or might not want to condition prices on the purchase history~\cite{acquisti2005conditioning}.

You run the local pharmacy and have sourced a large supply of X. Each day the same customer enters your store and appears interested in buying X. You are not sure about his willingness to pay. Let's assume he needs at most one X per day and he cannot store X appropriately so that the X he buys today, he cannot use tomorrow. Of course if you sold X successfully to him today, then you know a lower bound on his willingness to pay (WTP) tomorrow and could price X more accurately to extract more value from him it seems. However, if the buyer is forward-looking he might anticipate this and might be more reluctant to buy today, to not reveal information about his WTP for future interactions. 
For simplicity let's assume that the interaction happens in two periods (e.g. the day after tomorrow the customer will have forgotten about your pharmacy).
Suppose his willingness to pay is uniformly distributed on the unit interval and the same today as it is tomorrow.

In a Perfect Bayesian Equilibrium (PBE) of the game you sell X for $3/10$ today. If the customer buys X at the price of $3/10$, you double the price tomorrow. If the customer doesn't buy X today, you leave the price at $3/10$. The customer will buy X today if his WTP is at least $6/10$. Obviously he will buy X tomorrow if his willingness is at least the offered price ($6/10$ or $3/10$ depending on the situation). You will  believe that the customer's WTP is uniformly distributed between $6/10$ and $1$ in case you sell X today and uniformly distributed between $0$ and $6/10$ in case you don't sell X today. It is straightforward to verify that this is a PBE.

As you have a large supply of X you are of course wondering, whether it would be helpful for you to take X to forget that the customer who shows up today was already there yesterday. Why would that be? If you forget about the identity of the customer you treat him each day as the typical customer about whom you know nothing else. Thus, your best guess is that his value is uniformly distributed on the unit interval. In that case you would sell X at the same price $0.5$ each day. If you compare your revenue you will notice that you make $0.5=1/2*2*0.5+1/2*2*0$ on average if you are forgetful but only $0.45=3/10*(3/10+6/10)+4/10*(0+3/10)+3/10*0$ if you remember.

You are betrayed by your future self: It is in your best interest to extract a lot of rents once you have learned that the customer has a high willingness to pay. Thus, the customer is more careful in the present to hide his willingness to pay, even though it leads him to forgo an offer that he would otherwise accept. Your greedy future self hurts your bottom line today. 

You should take X to make more money from selling X.

\subsection*{Reputation and Amnesia}
Another class of examples where $X$ can benefit the player is when it helps him ignoring the other player's attempt to coerce him into taking certain actions by building a (threatening) reputation.  
\subsubsection*{Bargaining and Amnesia}
You go to the bazaar where the local carpet vendor offers you one of his magnificent rugs. The rug is advertised at a price of $\$10,000$. You are interested in the rug, you would in principle be willing to pay up to $\$12,000$ for it, but the offered price seems still excessive to you and you are convinced that the seller values it at $\$2,000$. However, you don't know whether the local carpet vendor is of the stubborn type who will stick to the advertised price regardless of what you do (you believe that he is stubborn with prob $\pi$) or of the rational type who might settle for a smaller offer after negotiation. The uncertainty is mutual. The carpet vendor does not know whether you are the stubborn type who will stick to an initial offer of $\$4,000$ (he believes that you're stubborn with prob $\pi$) or whether you might settle for a higher price after negotiation. After you both have made our initial offer, you either have a deal (in case of mutually acceptable offers) or you wait until one of you concedes and accepts the other's offer. This is the bargaining model with reputation of~\cite{abreu2000bargaining}. In the unique sequential equilibrium of the game you both choose a probabilistic strategy: You both initially mimic the stubborn type and offer to buy at $\$4,000$ resp.~to sell at $\$10,000$. Afterwards you (and symmetrically the vendor) randomly concede where the probability of conceding before time $0<t\leq T:=-3\log(\pi)/\beta$
 is
$$F(t)=1-e^{-\beta t/3}$$ where $0<\beta<1$ is the discount factor. A non-stubborn player will concede for sure by $T$ since he is now convinced that the other player is stubborn.

The equilibrium above a) is inefficient (because it takes time to come to an agreement) and b) relies on reputation. Reputation assumes memory,  put differently, reputation doesn't work on the forgetful. Can forgetting therefore restore efficiency?

\begin{figure}\label{fil}\centering
\includegraphics[scale=0.8]{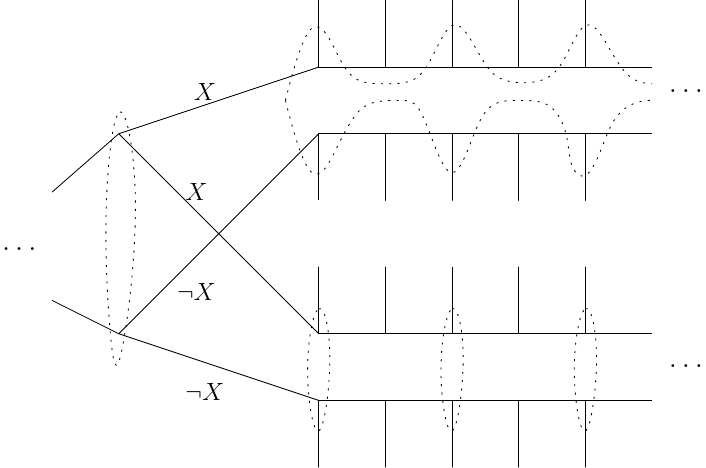}
\caption{Part of the game tree of a discrete version of the bargaining game: after the initial stages of nature choosing types and of the players making offer and counter-offer (not depicted), Player $1$ can take $X$ (not knowing Player 2's type). Taking $X$ makes him subsequently forget the number of times the concessions game has already been repeated (For readability, information sets of Player $2$ are not depicted).}
\end{figure}

Suppose you have access to X. If you take the pill, you remember the structure of the game, but you forget what has happened so far. See Figure~\ref{fil}. Thus, you don't know for how long you have already been at the bazaar. You only observe that the game hasn't ended yet. Note that if the vendor would employ the equilibrium strategy above even if you have access to X, your beliefs would look very different: If you take X and forget, your belief will be stationary and therefore your action will be stationary.

In the situation where you have access to X but the seller doesn't we can construct the following equilibrium: You take X so that you continuously forget what happens during the game, and you always offer to buy at $4000\$$. The seller concedes independently of what you do. If he observes you taking X he concludes that you don't concede because you choose to not remember what he does. When the seller observes you not-conceding and not taking the pill, he concludes that you're of the stubborn type and also concedes.
\subsubsection*{Mafia and Collective Amnesia}
Sometimes forgetting doesn't help individually but can help collectively.

The local mafia is engaged in extortion activity. Each day they extort a different business. The scheme is as follows: the extorted business can propose a high or low extortion amount. The mafia can accept the offer or kill the business owner. There is uncertainty about the type of the mafia. The business owner fears that the mafia could be of the reckless type which always kills the business owner when he offers only a small amount (and accepts if the amount is high). The business owner assigns probability $p$ to the mafia being reckless and probability $1-p$ to not being reckless. The non-reckless type of mafia behaves rationally, taking what they are offered to maximize their total payoff. The business owner observes what the mafia has done in the past, murders are reported in the local newspaper in grueling details and it is easy to figure out (although not provable in a court) that the mafia is behind it. It is also easy to figure out whether any business has paid a large sum to the mafia, gossip tends to spread fast in the business community. The payoffs in case the mafia is rational are 
\begin{center}
        \begin{game}{2}{2}
                  & High         & Low\\
     Accept       &$2,-2$       &$1,-1$\\
        Kill       &$0,-5$       &$0,-5$
        \end{game}
\end{center}

This is again a reputation game. It makes sense for the rational type mafia to act recklessly to build a reputation of recklessness. More specifically, they will kill the business owner whenever they observe a low offer and accept when they observes a high offer. If the business owner observes this behavior, they will make a high offer. More precisely, in the unique subgame perfect equilibrium, the (rational type) mafia plays "accept if high and kill if low" if so far no low offer has been accepted and ``always accept" if in some previous period a low offer has been accepted. Each business plays "high" if so far no acceptance of a low offer has been observed and "low" otherwise. On the equilibrium path, business owners make high offers which are accepted by the mafia.

Now imagine the business owners have access to X. Individually this does not make a difference.\footnote{If a business owner unilaterally takes $X$ to forget whether the mafia has acted recklessly in the past, this will not benefit them, as the mafia will still kill them when making a low offer (to build a reputation to future business owners).} But collectively, the business owners can profit from amnesia. Imagine the local business owners hold a meeting to which they invite the mafia to attend. At this meeting they collectively take X and this is observed by the mafia. After collective amnesia among business owners is established the game has a different subgame perfect equilibrium. In the equilibrium, the mafia "always accepts" and each business owner makes a low offer (now the business owners are not able to condition on the past).

\subsubsection*{Forward Induction and Amnesia}
The pill also helps to eliminate forward induction reasoning in games, again through the channel of amnesia.
You are the CEO  of a company and wonder whether your company should enter a market with one incumbent. After careful market analysis you come to the conclusion that there are two possible strategies that you and the incumbent could follow. After entry, the company can focus on a niche inside that market or try to capture a large chunk of the market. Likewise, the incumbent can choose, after observing market entry by the new competitor, to serve the niche or the main segment of the market. The choices which part of the market to serve (after entry) are made simultaneously. Let's assume that the payoffs are as follows:
%\begin{center}

 %       \begin{game}{2}{2}
  %                & main        & niche\\
   %    main      &$0,0$       &$3,1$\\
    %    niche     &$1,3$       &$0,0$
     %   \end{game}
      %  \end{center}
\begin{figure}[t]
    \centering
    \includegraphics[scale=0.7]{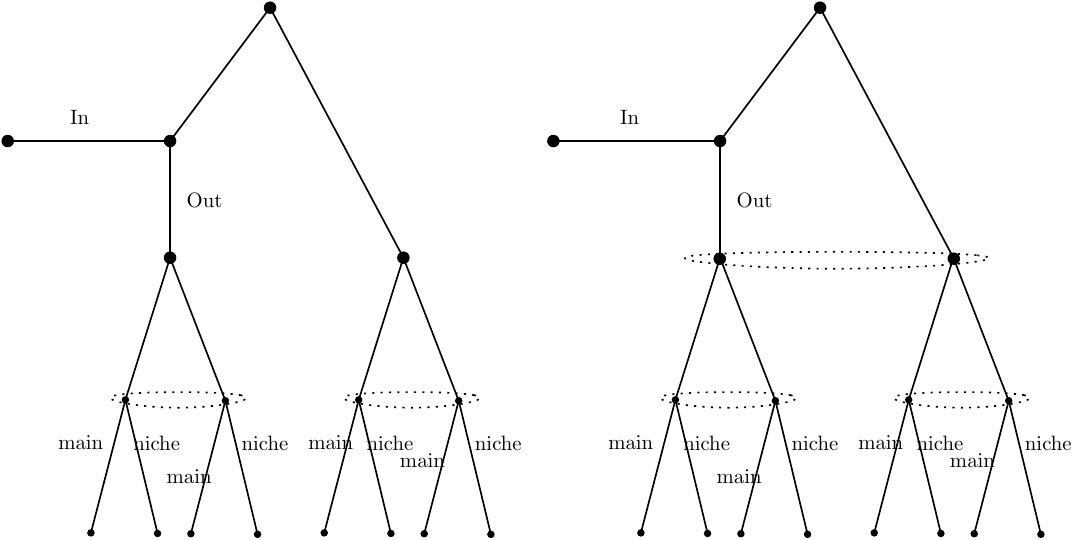}
    \caption{The original game (left): the entrant remembers after entrance that he has entered the market. The modified game (right): the entrant can choose to take $X$ after entrance which makes him forget whether he has entered the market.}
    \label{Untitled6}
\end{figure} 
Let's assume furthermore that there is a cost of entry of $2$ so that the payoff in case of no-entry are $2$ resp. $4$ (the incumbent can serve both market segments if you don't enter).
If both the entrant and incumbent choose the same strategy they will make no profit. If they choose different strategies, serving the main segment of the market is more lucrative than choosing the niche. In particular, entering the market (which is costly) is only worthwhile for the entrant, if the incumbent serves the niche market and the entrant gets the main market after entry. 
There are two pure strategy subgame perfect equilibria of this game: in the first you choose to enter the market and capture the main segment of the market after entry. The incumbent, after observing entry will go into the niche. In the second, the CEO chooses not to enter the market, because he expects that conditional on entry the incumbent will capture the main segment of the market and only leaves the niche for his company to capture.

The second equilibrium is often ruled out by forward induction reasoning: if you choose to enter the market you signal that you expect that the incumbent goes into the niche, otherwise you wouldn't enter in the first place. Thus, the incumbent infers that the entrant whenever he enters will go for the main market and hence best replies by choosing the niche. In other words, the beliefs of the incumbent's CEO in the second equilibrium are not really compatible with his observations. He acts as if he would forget... This seems implausible.

Now suppose the incumbent's CEO has access to X. Forgetting now is very plausible. More precisely,  we modify the game as follows:
Nature randomly chooses at the beginning whether the game is a game of entry as described above or the game is one where the entrant is already in the market (the entry phase is not part of the game). Without the pill, the incumbent knows at which information set he is when making a decision and therefore knows whether they are in the entry game or in the game where the competitor has been in the market forever. Thus, in the subgame with entry the forward induction reasoning applies. If the incumbent has access to the pill and nature chooses the game without entry with sufficiently high probability, then forward induction reasoning has no power. We have an equilibrium where the incumbent gets the main segment and the entrant the niche and the entrant chooses not to enter the market.

%Over here the interesting thing is that delegation is different from taking pill X. So: amnisia is better than blockchain.

\subsection*{One time use of information}
Another powerful application of X is the enforcement of one time (or limited number of times) use of information. This is useful in variety of situations where sharing information can lead to a better outcome in a specific context, but agents might fear that the information is used in later situations to their detriment. 
\subsubsection*{Information sharing in Bargaining}
 Bilateral bargaining is a classical situation where sharing information can increase efficiency because it makes an agreement more likely. However, the parties might not want the information to be used outside of the negotiation process. Thus, without the ability to conditionally share information, they might be less willing to share information during the negotiation process leading to a higher likelihood of bargaining failure. 

\subsubsection*{Corporate} In corporate negotiations, non--disclosure agreements (NDAs) are signed to prevent leak of information and dictates that the information should not be used in a way that impacts the informants' behavior outside of the negotiation, it's essentially conditional recall but works at the natural language level. With pill X, there would not be a need for NDAs as negotiators would only be able to temporarily access private information and use it in the boardroom. 

\subsubsection*{War} War should theoretically not occur in equilibrium unless there is incomplete information or a misalignment of preferences. In a scenario with complete information and alignment between decision-makers and those executing the war, countries would logically exchange resources rather than engage in conflict, as the outcome of the war would be predictable in advance. However, in reality, predicting the outcome of a war is challenging without full knowledge of the opposing side's military strength. Revealing such information—such as troop numbers or advanced weaponry—could compromise strategic advantages, making it unwise to disclose these details. This secrecy often results in mismatched expectations, leading both parties to a lose-lose situation by choosing war over negotiation.\footnote{In a hypothetical world with complete information, war might be avoided entirely. Jorge Luis Borges provides an apocryphal example (``Libros y autores extranjeros", book review of Erich Ludendorff's ``Der totale Krieg."): ``In 15th century Italy, war had reached a perfection that many would call ridiculous. Once the armies were assembled, the generals compared the numbers, strength, and positioning of their troops and decided who among them must suffer defeat. Chance and bloodshed were eliminated." So essentially, generals can mimic a real war by agreeing to all present a sample of their military at the same place at the same time, and then they can negotiate based on the partially-revealed information. A historical variant of this concept occurred in the Battle of the 300 Champions, where two Greek states each sent 300 elite soldiers to fight on behalf of their armies, avoiding full-scale conflict.}

With pill X and conditional recall, it's possible to implement a mechanism where each country send their general to the other country's military base, review all of the firearms, and then negotiate the result of the war. This prevents either general from launching a followup attack after learning the other general's tactics and using that knowledge as an unfair advantage. Without the ability to conditionally recall, generals would not choose to reveal secret weapons or their surprise attack tactics because the other party may choose to dishonor the negotiation after learning those and think they can win the war with that knowledge. This would then lead to both generals withholding information and therefore make the negotiation not honored and therefore ending in a total war.

% In this situation, delegation with conditional recall is helpful because it allows you to credibly signal ignorance of your own map such that you can have a credible commitment over your asset to do threats and agreements.

% Moreover, the design of such a contract would involve you committing to credibly only use your information with that cartel and not doing anything else (physical presence, TEE is also using physical presence to enforce some magic in the digital/idea land).

\subsubsection*{Account delegation}

You’re planning a vacation and want to disconnect completely from the stress and noise of the internet. Knowing you wouldn’t be able to resist the temptation of using your devices, you decide to leave them at your workplace. But a thought nags at you: What if something happens to my parents, and the hospital needs to call me? You try to convince yourself to bring the devices along, using this as justification.

Your friend, concerned about you sabotaging your trip, offers a solution: they’ll monitor your calls, social media, and emails while you’re away. If anything urgent arises, they’ll handle it for you. Still, the idea makes you uneasy. What if they see messages you're embarrassed about? Despite your friend’s offer, your reluctance and anxiety lead to a predictable outcome—you spend the entire vacation doom-scrolling through Twitter.

Now, imagine your friend had access to X. The pill would let them filter and respond to only critical, family-related messages without exposing your private conversations. Similarly, you could think of this as using an LLM-powered email filter: raw data is processed, sensitive information is cleansed, and only essential details are passed along, all under a limited-use token. These types of applications are examples of ``one-time programs"\cite{OneTime}.

% \textbf{Darkleaks}

% but essentially its limited to either part of the passage for free, or sharing metadata only.... it can give a ZK Proof of provenance, but that's not sufficient because that's not revealing privileged information just for use during the negotiation, besides what's fully revealed
\subsubsection*{Matching} 

In matching markets, it’s tempting to misrepresent your traits to improve your matches. Over time, if everyone starts exaggerating or lying about their qualities, the market risks collapsing under the same principles seen in the "Market for Lemons" example. Take a dating app, for instance: it’s easy to claim you have desirable traits like a love for reading, a passion for traveling, or a good sense of humor. Verifying such claims, however, is costly and invasive. For example, proving someone’s love of books might require examining their daily schedule or purchase history, while verifying shared tastes in movies could involve revealing a Netflix recommendation page. Understandably, most people are uncomfortable exposing such personal data without a certain level of commitment, leaving these traits unverifiable and the market susceptible to adverse selection.

X offers a potential solution. It allows users to safely share and verify sensitive information without overexposing themselves. For instance, a trusted system could review a person’s reading habits or entertainment preferences, confirm their authenticity, and provide a simple ``stamp of approval" for specific traits. Adverse selection typically arises when verifying private information is too costly, invasive, or subjective. By enabling efficient and secure verification,  X expands the range of traits that can be confidently validated, stabilizing the market and reducing the incentive to misrepresent oneself.

\subsubsection*{Spoilers} In the world of entertainment, spoilers can significantly impact one's enjoyment of a movie, book, or TV show. Currently, people rely on social norms and warnings to avoid spoilers, but these methods are imperfect. With pill X, one could safely engage in discussions about a piece of media without risking exposure to unwanted information. For instance, a fan could attend a Question and Answer session with the creator of a TV series, gaining insights into the creative process without accidentally learning plot points they haven't reached yet. This would allow for more open and in-depth discussions within fan communities without the constant fear of spoilers, and we are already seeing real-life\footnote{https://chat.openai.com/g/g-Usa50OPO4-spoiler} implementations of this.

\subsubsection*{Acquisitions} In the business world, information about potential acquisitions can have significant unintended consequences. When a large company expresses interest in acquiring a smaller firm, the mere knowledge of this interest can cause the target company's stock price to spike, potentially making the acquisition more expensive or even unfeasible. ``If oil company wants to buy your house, there is oil underneath." With pill X, negotiators could explore potential acquisitions without this information leaking to the market. They could conditionally recall the details of their discussions only when in specific negotiation settings, preventing premature market reactions and allowing for more efficient and discreet business dealings.

\subsubsection*{Juries \& De-Biasing}
You are called to jury duty in a murder trial. As part of the jury, your responsibility is to determine the facts of the case based on the evidence and testimony presented. However, not all evidence is admissible in court, even if it could potentially aid in uncovering the truth. For instance, evidence obtained illegally is excluded to ensure the integrity of the judicial process.

Now imagine a piece of inadmissible evidence comes to your attention. This information, while potentially helpful, might taint your judgment and conflict with your duty to base your decision solely on the admissible evidence. It’s natural to want to uncover the unconditional truth,\footnote{The evidence on whether juries are more likely to acquit if they have seen inadmissible evidence that points to the guilt of the defendant is mixed, but seems correlated with the severity of the prosecuted crime. In experiments, jurors faced with a hypothetical case, have a bias (for acquitting) if faced with exonerating evidence, but not if faced with other evidence.} yet your role requires you to focus on the most likely truth implied by the legally permissible evidence.

Here, X could help resolve this conflict. By taking it, you could focus only on the admissible evidence, filtering out the influence of irrelevant or prohibited information.
 There are other applications of de-biasing (or biasing): an LLM which has been trained on data that is biased (or was obtained illegally, e.g. in breach of intellectual property) or human agents with learned biases can un-learn their biases. This has various applications from fairer job recruitment, credit scoring, medical
diagnosis systems to better recommendation systems. Of course, the same technology can also be used for malicious intentions by introducing biases through withholding relevant information.
% \subsubsection*{Recommendation System}
\section{Conclusion}
We have explored the strategic advantages of credible, conditional forgetting across various contexts, illustrating how the hypothetical Pill X  can enhance efficiency in many strategic interactions with potentially far-reaching implications. While the ability of credibly forgetting is just a thought experiment for human agents, it is a realistic possibility for artificial agents, as we sketched in the introduction. Even more so, several of the applications, such as the account delegation example are already implementable or implemented in reality through a combination of LLMs and account encumbrance by cryptographic means such as TEEs. Although human agents may never possess the ability to credibly and conditionally forget, they can still benefit from delegating decision-making to artificial agents equipped with this capability. As we’ve shown, this approach can serve as a cheap substitute to traditional commitment devices that rely on the legal system. More intriguingly, it also opens up entirely new possibilities that go beyond the scope of conventional mechanisms, offering exciting opportunities for innovation in mechanisms design.
%The world can become more incoherent.

%Possible problem is that players remember elements of the past.
%Because decisions nodes can be distinguished by the set of available actions. 

%Talk about various movies.

%This is linear types and one-time-programs in reality. 
\section{Acknowledgments}
We thank Quintus Kilbourn, Sarah Allen, Leo Arias, Akaki Mamageishvili and Matt Stephenson and attendees of our Devcon talk for feedback on an earlier version of this paper. 
\bibliographystyle{ACM-Reference-Format} 
\bibliography{arxiv.bib}

\end{document}